\documentclass[preprint]{aastex}
\usepackage{graphicx}        % For eps figures, newer & more powerfull
\usepackage{color}           % For color text: \color command
\usepackage{breakurl}        % For breaking URLs easily trough lines
\slugcomment{}
\begin{document}

\shortauthors{Inceoglu et al.}
\shorttitle{On the current solar magnetic activity}
\title{On the current solar magnetic activity in the light of its behaviour during the Holocene}

%\author[addressref={aff1,aff2,aff3},corref,email={e-mail.a@mail.com}]{\inits{F.I.}\fnm{F. Name}~\lnm{Author-a}}%\sep
%\author[addressref=aff1,email={e-mail.b@mail.com}]{\inits{F.}\fnm{Fisrt}~\lnm{Author-b}}%\sep
%\author[corref,email={e-mail.c@mail.com}]{\inits{S.}\fnm{Second}~\lnm{Author-c}}%\sep
%\author{\inits{T.}\fnm{Third}~\lnm{Author-x}}

\author{F.~Inceoglu}
\affil{Stellar Astrophysics Centre, Department of Physics and Astronomy, Aarhus University, Ny Munkegade 120, DK-8000 Aarhus C, Denmark}
\affil{Department of Geoscience, Aarhus University, H{\o}egh-Guldbergs Gade 2, DK-8000 Aarhus C, Denmark}
\email{fadil@phys.au.dk}

\author{R.~Simoniello}
\affil{Laboratoire AIM, CEA/DSM-CNRS-Universit{\'e} Paris Diderot, IRFU/SAp, Centre de Saclay, F-91191, Gif-sur-Yvette, France}
\email{Rosaria.Simoniello@unige.ch}

\author{M.~F.~Knudsen}
\affil{Department of Geoscience, Aarhus University, H{\o}egh-Guldbergs Gade 2, DK-8000 Aarhus C, Denmark}

\author{C.~Karoff}
\affil{Stellar Astrophysics Centre, Department of Physics and Astronomy, Aarhus University, Ny Munkegade 120, DK-8000 Aarhus C, Denmark}
\affil{Department of Geoscience, Aarhus University, H{\o}egh-Guldbergs Gade 2, DK-8000 Aarhus C, Denmark}

\author{J.~Olsen}
\affil{AMS, $^{14}$C Dating Centre, Department of Physics, Aarhus University, Ny Munkegade 120, DK-8000 Aarhus C, Denmark}

\affil{S.~Turck-Chi\`eze}
\affil{Laboratoire AIM, CEA/DSM-CNRS-Universit{\'e} Paris Diderot, IRFU/SAp, Centre de Saclay, F-91191, Gif-sur-Yvette, France}

\begin{abstract}

Solar modulation potential (SMP) reconstructions based on cosmogenic nuclide records reflect changes in the open solar magnetic field and can therefore help us obtain information on the behaviour of the open solar magnetic field over the Holocene period. We aim at comparing the Sun's large-scale magnetic field behaviour over the last three solar cycles with variations in the SMP reconstruction through the Holocene epoch. To achieve these objectives, we use the IntCal13 $^{14}$C data to investigate distinct patterns in the occurrences of grand minima and maxima during the Holocene period. We then check whether these patterns might mimic the recent solar magnetic activity by investigating the evolution of the energy in the Sun's large-scale dipolar magnetic field using the Wilcox Solar Observatory data. The cosmogenic radionuclide data analysis shows that $\sim$71\% of grand maxima during the period from 6600 BC to 1650 AD were followed by a grand minimum. The occurrence characteristics of grand maxima and minima are consistent with the scenario in which the dynamical non-linearity induced by the Lorentz force leads the Sun to act as a relaxation oscillator. This finding implies that the probability for these events to occur is non-uniformly distributed in time, as there is a memory in their driving mechanism, which can be identified via the back reaction of the Lorentz force.
  
\end{abstract}
\keywords{Sun: activity -- Sun: dynamo -- Solar-terrestrial relations}

\section{Introduction}

Sunspot observations since 1610 have revealed that the Sun has an 11-year cyclic magnetic activity, which is modulated on longer time-scales in association with grand minima, such as the Maunder Minimum (1645 -- 1715), when sunspots vanished almost completely from the solar photosphere, and with grand maxima, when the magnetic activity levels are enhanced, such as the Modern Maximum\footnote{This maximum corresponds to the period of relatively high solar activity which began with solar cycle 15 in 1914. It reached a maximum in solar cycle 19 during the late 1950s.}\citep{2007A&A...471..301U}. \citet{1993A&A...276..549R} showed that during the recovery phase of the Maunder Minimum, the sunspot activity displayed a north-south asymmetry where sunspots were mainly observed in the southern solar hemisphere, while sunspots were dominant in the northern solar hemisphere during the Dalton Minimum \citep{2009ApJ...700L.154U}. Furthermore, using historical reconstructions of the butterfly diagram for the period of 1749--1796, \citet{2009SoPh..255..143A} provided lines of evidence that the quadrupolar component of the Sun's large-scale magnetic field played a dominant role compared to the dipolar one. Following this period, the north-south asymmetry ceased and the Sun's large-scale magnetic field became dipolar again when Hale's polarity cycle was obeyed \citep{1997A&A...322.1007T}. 

Information on solar variations prior to 1610 relies on past production rates of cosmogenic nuclides, such as $^{10}$Be and $^{14}$C \citep{2013LRSP...10....1U}. The production rates of cosmogenic nuclides depend on the intensity with which cosmic rays impinge on the Earth's atmosphere \citep{Dunai2010}. However, before reaching the Earth, cosmic-ray particles have to travel through the heliosphere \citep{2013LRSP...10....3P}, where they become modulated by the heliospheric magnetic field, which is a component of the Sun's large-scale magnetic field \citep{2002GeoRL..29.2224L,2004SoPh..224...21W}. The production rates of cosmogenic radionuclides are inversely correlated with solar magnetic activity and the geomagnetic field intensity. This inverse correlation is caused by the non-linear shielding effect of the solar magnetic field and the geomagnetic dipole field \citep{1967HDP....46..551L}. Therefore, the flux of Galactic Cosmic Ray (GCR) particles reaching Earth is inversely proportional to, but out of phase with, the 11-year solar cycle. Temporal variations in the GCR intensity display a distinct 11-year periodicity owing to solar modulation of GCRs by the heliospheric magnetic field component of the large-scale field of the Sun in the heliosphere \citep{2002GeoRL..29.2224L,2013LRSP...10....3P}. 

Earlier studies based on past production rates of cosmogenic nuclides have shown that the Maunder Minimum and the Modern Maximum were not the only quiescent and enhanced activity periods observed over the last $\sim$10.000 years. \citet{2015A&A...577A..20I} suggest that during the period from 6600 BC to 1650 AD, the Sun experienced 32 grand minima and 21 grand maxima, whereas \citet{2007A&A...471..301U} claimed that the Sun underwent 27 grand minima and 19 grand maxima since 8500 BC based on their sunspot number reconstructions.

Some specific aspects of the long term variations in amplitude and parity\footnote{changes in the symmetry in the field topology.} can be reproduced by the current solar/stellar dynamo models. For example, grand minima are seen as quiescent intervals of activity that interrupt periods of moderate activity levels. There are at least two ways to reproduce these intermittent periods in solar dynamos; via sudden changes in the governing parameters of the solar dynamo \citep{2008SoPh..250..221M} or via the back-reaction of the Lorentz force on the velocity field, acting as a dynamical non-linearity \citep{1995MNRAS.273.1150T,1996A&A...307L..21T,1997A&A...322.1007T}. In the latter scenario, the back reaction of the Lorentz force on the velocity flow arising from the magnetic field generated by the solar dynamo is taken into account, which tends to oppose the driving fluid motions through velocity fluctuations, the so-called Malkus-Proctor effect \citep{1975JFM....67..417M}. This mechanism has also been shown to cause amplitude and parity modulation of the Sun's large-scale magnetic field \citep{1997A&A...322.1007T}. Especially, when the magnetic field is particularly strong, the Sun might act as a relaxation oscillator, which forces the solar magnetic activity into a minimum phase after a period of strong activity \citep{1997A&A...322.1007T,2008GeoRL..3520109A}. 

In this study, we look for observational evidences in support of the scenario explained above, whether such a pattern indeed is present in the data as it was suggested that grand minima and maxima occur when the solar dynamo enters a special state. This implies that the occurrence and nature of grand minima and maxima do not reflect purely random processes without a memory, but there is a memory in the driving process \citep{2015A&A...577A..20I}. The findings might help us interpret the current solar dynamics as observations of solar activity structures, \textit{e. g.} sunspots, coronal mass ejections, solar winds, may imply that a period of reduced magnetic activity after the Modern Maximum could have already started during the descending phase of solar cycle 23 \citep{2004ApJ...601.1136D,2011GeoRL..3819106O,2014ApJ...784L..27W}. We, therefore, investigate the solar dynamics during the time period from 6600 BC to 1650 AD attempting to identify regular patterns of occurrence in grand minima and maxima. This categorisation may allow us to classify the recent solar dynamics.

%__________________________________________________________________
\section{Characterisation of the solar dynamics during the Holocene}

\subsection{The cosmogenic radionuclide data}

In this study, we use the grand solar minima and maxima periods identified in \citet{2015A&A...577A..20I}, who used the GRIP $^{10}$Be and IntCal13 $^{14}$C records for the overlapping period from 6600 BC to 1650 AD to reconstruct the solar modulation potentials (SMPs) and they detected 32 grand minima and 21 grand maxima based on three criteria, \textit{i. e.}, (i) the amplitude of low and high activity periods have to be above and below threshold values, respectively, which are calculated based on the distributions of overlapping peaks and dips, (ii) the durations of these periods have to be longer than two solar cycles, and (iii) they have to occur simultaneously in the two reconstructions based on the GRIP $^{10}$Be and IntCal13 $^{14}$C records \citep[see][for details]{2015A&A...577A..20I}. Using such a method provides a more robust identification of grand minima and grand maxima, since the geochemical behaviour of these two radionuclides are different and their deposition rates can be affected by different processes in the Earth system \citep{2013CliPa...9.1879R}. In addition to the list given in \citet{2015A&A...577A..20I}, we also included the Maunder Minimum (1645 -- 1715 AD.) to the list of grand minima, which could not be included in that study because of the time span of the GRIP $^{10}$Be record, which extends up to 1650 AD. It's worth noting that all the numerical results presented here are based on the SMP reconstruction calculated by \citet{2015A&A...577A..20I} using the IntCal13 $^{14}$C record, which extends up to 1950 AD. However, because of the Suess effect, which has caused a significant decrease in the $^{14}$C/$^{12}$C ratio as a consequence of admixture of large amounts of fossil carbon into the atmosphere after the industrial revolution \citep{1955Sci...122..415S}, we disregarded data after 1820.

\subsection{Distribution of Grand Minima and Grand Maxima}

\begin{figure}
\resizebox{12cm}{!}
{\includegraphics{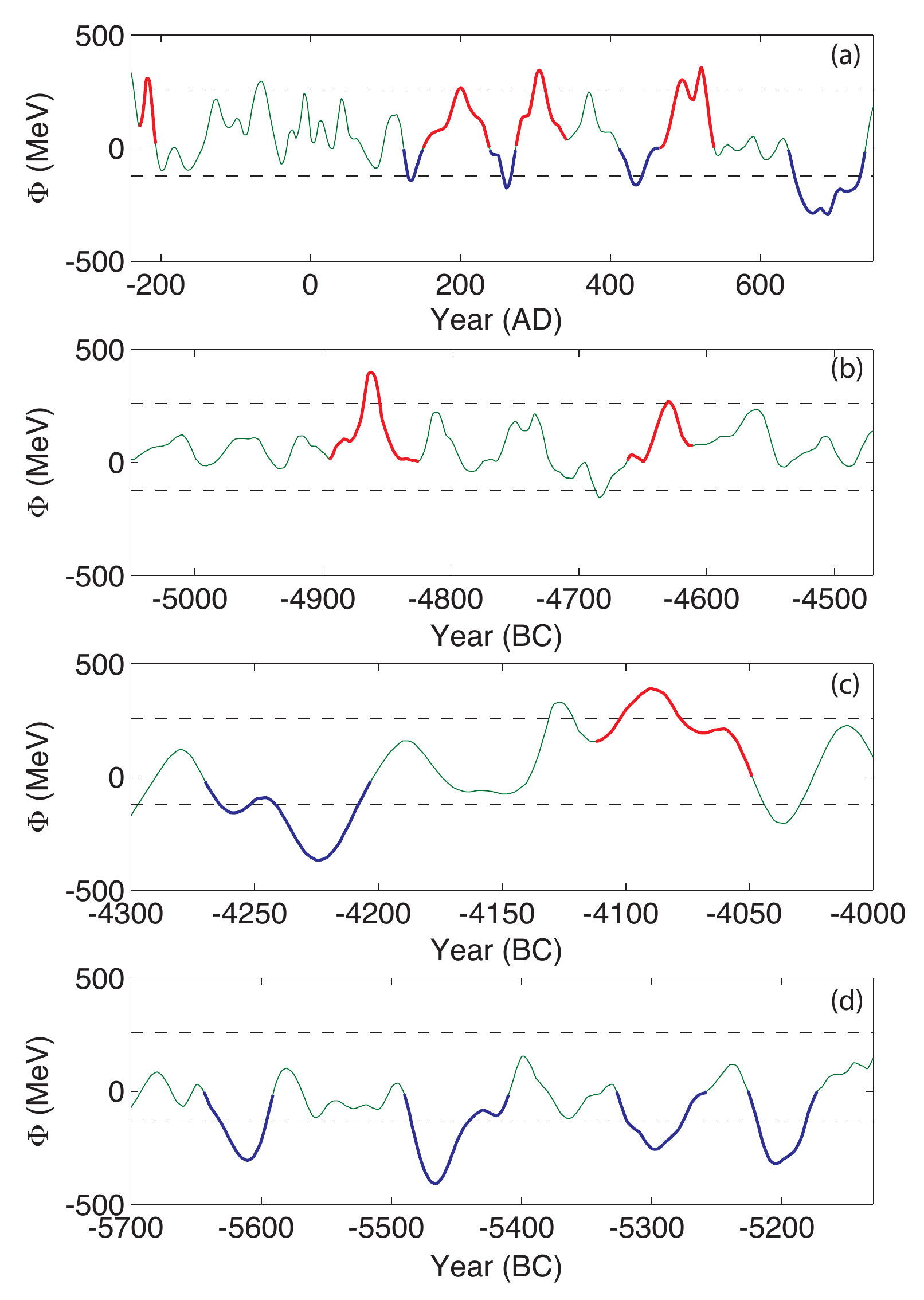}}
\caption{The categorisation of grand minima (blue) and grand maxima (red) in four distinct groups: (a) grand maxima followed by a grand minimum, (b) grand maxima followed by another grand maximum after a period in a moderate activity level (green), (c) grand minima followed by a grand maximum, and (d) grand minima follow a grand minimum.}
\label{fig:MaxMinsub}
\end{figure}

We investigated statistical occurrence characteristics of grand maxima and minima and found that 15 out of 21 grand maxima are followed by a grand minimum (Figure~\ref{fig:MaxMinsub}a) either immediately after or after spending some time in a moderate activity period, meaning that $\sim$71$\%$ of grand maxima are followed by a grand minimum and almost 47\% of the grand minima follow a grand maximum. Table~\ref{tab:GrandMinima} lists these events together with their duration.

\begin{table}
\caption{The center times and durations of the grand minima, which follow a grand maximum. Minus signs before the dates represent years BC.}
\begin{tabular}{cc}
\hline \hline
Center time & Duration  \\  \hline
1695& 84\\
1450&167\\
690&102\\
434&54\\
261&34\\
133&25\\
-348&107\\
-1491&53\\
-2461&74\\
-2874&106\\
-3080&28\\
-3330&134\\
-3695&30\\
-4322&72\\
-5713 &41\\
\hline
\end{tabular}
\label{tab:GrandMinima}
\end{table}

There are also six grand maxima events (29$\%$) that are not followed by a grand minimum and 17 grand minima events (53$\%$) that are not following a grand maximum. Figure~\ref{fig:MaxMinsub} shows examples of the four different categorisations based on identifications of grand minima and maxima from the SMP reconstruction based on the $^{14}$C data: (i) an example of grand maxima followed by grand minima either immediately after or after spending some time in a moderate activity level (Figure~\ref{fig:MaxMinsub}a), (ii) six grand maxima events that are followed by another grand maximum after spending some time in a moderate activity level (one example shown in Figure~\ref{fig:MaxMinsub}b), (iii) six grand minima that are followed by a grand maximum after a period of moderate activity (an example shown in Figure~\ref{fig:MaxMinsub}c), and (iv) 11 grand minima, which follow another grand minimum immediately after or after a period of moderate activity (an example shown in Figure~\ref{fig:MaxMinsub}d). We must note that we only analyse the statistical recurrence of grand minima and maxima based on the four categorisations above, therefore any intrinsic time-scale of the solar dynamo action is not considered here. It's also worth noting that the peak around 70 BC (Figure~\ref{fig:MaxMinsub}a) is not considered as a grand maximum, since it doesn't occur simultaneously in the SMP reconstruction based on the GRIP $^{10}$Be record \citep[see][for details]{2015A&A...577A..20I}.

To test the statistical significance of the fraction of the grand maxima that are followed by a grand minimum identified using the criteria defined in \citet{2015A&A...577A..20I} against the null hypothesis of the purely random occurrences of grand solar minima and maxima, we use the Monte Carlo test. We first generated a thousand red noise datasets, which have the same mean and standard deviation values as the real data. To identify grand minima and maxima periods in the simulated data sets, we apply the selection criteria given in Section 2.1 \citep[see][for details]{2015A&A...577A..20I}, except from the selection criterion requiring that low- and high-activity periods, respectively, have to overlap in two datasets. This criterion is excluded to achieve a sufficient number of grand minima and maxima in the simulated data. Inclusion of this criterion would result in very small numbers of grand minima and maxima that are overlapping in two simulated data sets. Therefore, to pursue the statistical significance analysis based on these criteria, we first generate a red noise data set. Subsequently, we determine the grand minima and maxima periods for each resampled data set using the threshold values found in \citet{2015A&A...577A..20I}. Additionally, to make a better comparison of the results of the simulated data with the results of the underlying data, we also exclude the same selection criterion for the real data set. The results show that $\sim$72\% of grand maxima are followed by a grand minimum (21 grand maxima out of 29) in the $^{14}$C based SMP reconstruction.

\begin{figure}
\begin{center}
{\includegraphics[width=3.5in]{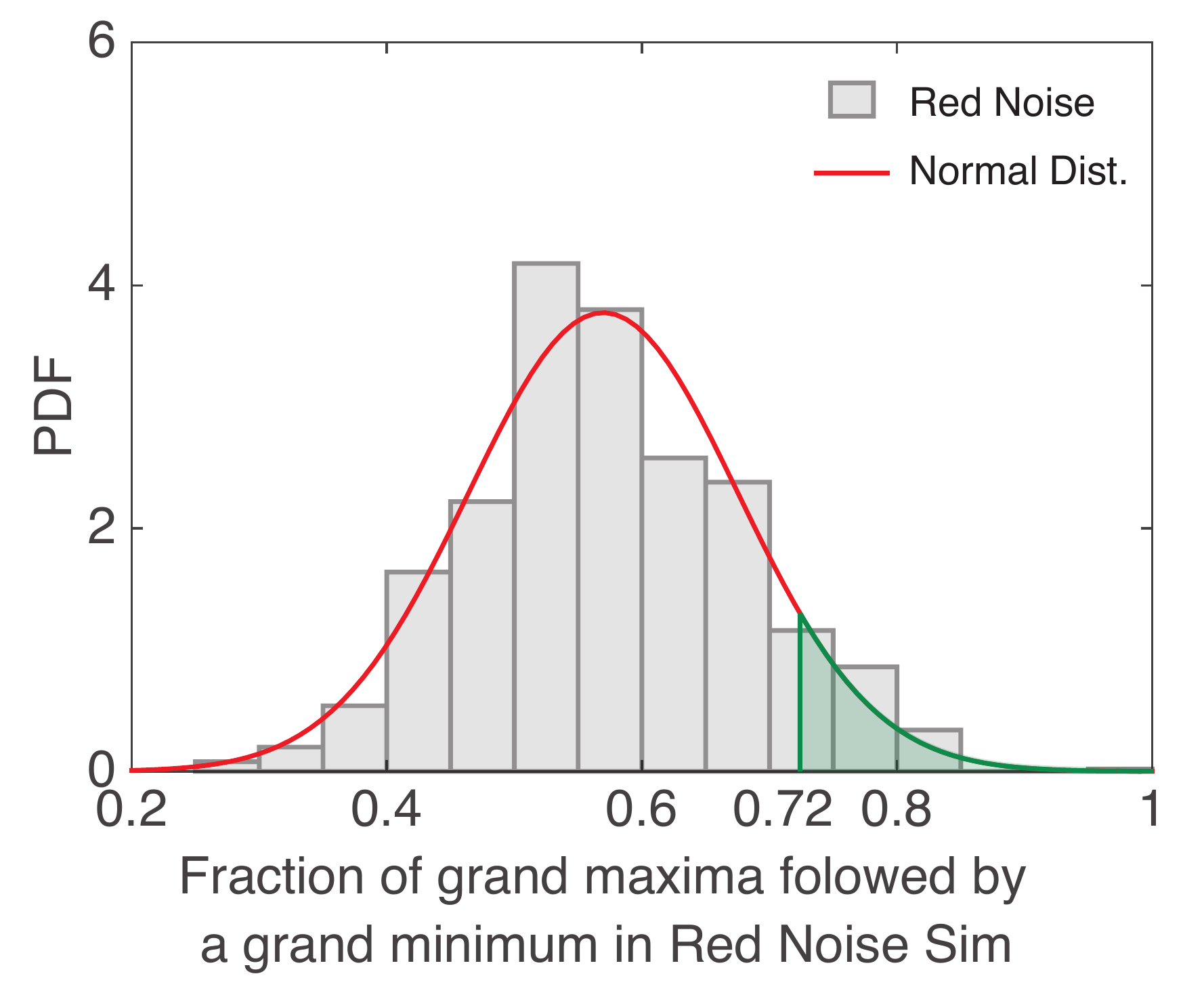}}
\caption{The probability density function (PDF) of the fraction of grand maxima that are followed by a grand minimum identified in each red noise data (grey). The green shaded area on the right side of 0.72 (see text) is 0.07. The red line shows the Gaussian distribution fitted to the PDF.} 
\label{fig:StatSig}
\end{center}
\end{figure}

The resulting probability density function (PDF) for the fraction of grand maxima that are followed by a grand minimum found in the simulated red noise data is shown in Figure~\ref{fig:StatSig}. For the red noise data (grey), the fraction of grand maxima that are followed by a grand minimum is best represented by a Gaussian distribution (Figure~\ref{fig:StatSig}). The distribution implies that the $\sim$72\% of grand maxima that are followed by a grand minimum in the real data set ($\sim$71\% when the selection criterion of overlap included) might be a pattern inherent in the real data. The area on the right side of the value of $\sim$72\% is $\sim$0.07 (Figure~\ref{fig:StatSig}), implying that the observation of 72\% of grand maxima being followed by a grand minimum based on the $^{14}$C based SMP reconstruction is marginally significant at $p=\sim0.93$ level. 

\subsection{Grand Maxima followed by a Grand Minimum}
 
\begin{figure}
\resizebox{\hsize}{!}
{\includegraphics{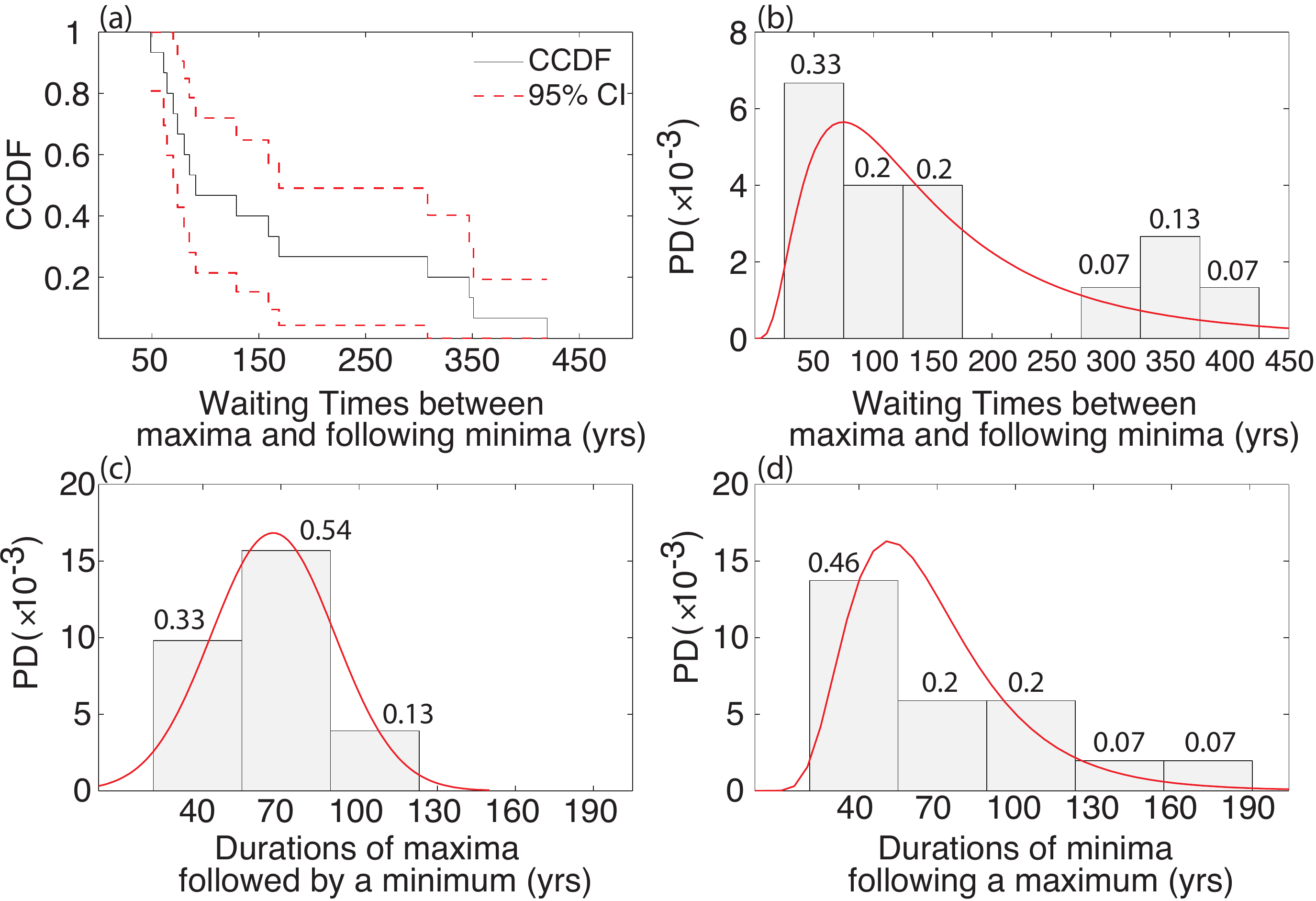}}
\caption{(a) the complementary cumulative distribution function (CCDF) of waiting times between subsequent maxima and minima; (b) the probability densities (PDs) of waiting times with a log-normal fit (red line); (c) the PDs of the durations of maxima that are followed by minima with unimodal gaussian fit (red line); and (d) the PDs of the durations of minima that are following maxima with a log-normal fit (red line) based on $^{14}$C data. In panels b, c, and d, also shown the occurrence probabilities of the events as a function of the waiting-times above the bars (see text).}
\label{fig:MaxFBMin}
\end{figure}

The 15 grand maxima events, which are followed by a grand minimum, may represent the current solar dynamics, as the Modern Maximum is thought to have ended with solar cycle 23 \citep{2001SoPh..203..179R,2013LRSP...10....1U}. Therefore, we calculated the waiting times between subsequent grand maxima and grand minima in $^{14}$C based SMP reconstruction. Waiting times can be considered as intervals between subsequent peaks or dips in activity \citep{1998ApJ...509..448W}. Figure~\ref{fig:MaxFBMin}a shows the complementary cumulative distribution function (CCDF) of the waiting times between the peak and the dip times of subsequent grand maxima and minima together with the distributions of durations of maxima periods, which are followed by a grand minimum, and durations of grand minima periods, which follow a maximum. The CCDF is defined as the probability that an event "X" with a certain probability distribution will be found at a value more than or equal to "x". The mathematical formulation is shown below;

\begin{equation}
{\rm P~(X \ge x)}=1-\int_{-\infty}^{~x}{\rm p(x)~dx}=\int_{~x}^{~\infty}{\rm p(x)~dx}
\label{eq:CCDF}
\end{equation}

\noindent where p(x) denotes the probability density (PD). The relative probability to have a grand minimum state after a grand maximum decreases with time (Figure~\ref{fig:MaxFBMin}a). 

We then tested whether the distributions shown in Figure~\ref{fig:MaxFBMin}b, c and d are best represented by unimodal Gaussian (normal), lognormal, or  bimodal Gaussian distributions by comparing the Akaike Information Criterion (AIC) and the Bayesian Information Criterion (BIC) values of the three fits. \citet{Schwarz1978} suggested the BIC as an alternative to the AIC, previously suggested by \citet{Akaike1974}. The BIC and AIC are calculated using the following equations \citep{Akaike1974,Schwarz1978};

\begin{eqnarray}
&{\rm AIC}=-2\times \ln L + 2k, \label{subeq1} \\
&{\rm BIC}=-2\times \ln L + k\times \ln(n) \label{subeq2} 
\end{eqnarray}

\noindent where n and k denote the sample size and number of estimated free parameters included in the model, respectively. The likelihood of the model, shown by the term $L$, reflects the overall fit of the model \citep{BurnhamAnders2002}. Under the assumption that the residuals between the observed and the modelled values are Gaussian distributed, $L$ is calculated as follows \citep{2013MNRAS.430.2313C};

\begin{equation}
L = \prod_{i=1}^{N}\frac{1}{\sqrt{2\pi}\sigma_i}\exp\left[-\frac{1}{2} \left(\frac{\Delta_i}{\sigma_i}\right)^2\right]
\end{equation}

\noindent where N is the total number of data points, $\Delta_{i}$ represent the residuals between observed and modelled quantities and $\sigma_i$ is the uncertainty in observations. Even though the BIC and AIC are extensions of the maximum likelihood principle, they exhibit differences in some aspects; the BIC is generally used when the main aim is to build a model that describes the distribution of the data, whereas the AIC is used for more predictive aspects \citep{NeatCav2012}. According to \citet{KassRaf95}, if the difference between the BIC values of two distributions (BIC$_{A}$-BIC$_{B}$) is  less than 2, then the likelihood of model "A" is comparable to that of model "B". For differences between 3 and 7, there is a strong evidence that model "B" is more likely to represent the distribution better, whereas model "A" has considerably less support. When the difference between BIC$_{A}$ and BIC$_{B}$ is larger than 10, it is decisive that model "B" is the most likely model that best represents the data, while model "A" is very unlikely. The same rule is also valid for the AIC values \citep{BurnhamAnders2002}. 

\begin{table}
\caption{The Bayesian (BIC) and Akaike (AIC) information criterion values for unimodal Gaussian, lognormal and bimodal Gaussian fits to the waiting times between maxima and their following minima, durations of grand maxima, which are followed by a grand minimum, and grand minima, which are following grand maximum.}
\begin{tabular}{lcccccc}
\hline \hline
						& \multicolumn{2}{c}{Unimodal Gaussian}							&\multicolumn{2}{c}{Bimodal Gaussian} 	& \multicolumn{2}{c}{lognormal}		\\ \hline
				& BIC	& AIC								&BIC 	 & AIC					&BIC 	 & AIC				\\ \hline
Waiting Time		&192		&191									&182		&179						&183		&181					\\
G. Max.			&142		& 140								&146		 & 142					&146		&144					\\ 
G. Min.			&159		& 158								&161		 & 157					&156		&155					\\ 
 \hline
\end{tabular}
\label{tab:Durations}
\end{table}

The calculated BIC and AIC values (Table~\ref{tab:Durations}) show that the waiting times between maxima and their following grand minima can be represented either by a log-normal or a bimodal Gaussian distribution, but it is not a decisive suggestion. The durations of grand maxima periods are best represented by a normal distribution with a mean duration of $\sim$65 years, while a lognormal distribution with a mean duration of $\sim$75 years represents the distribution of grand minima durations better (Table~\ref{tab:Durations}). It must be noted that the significance of these findings are not decisive and may be caused by the sample size of the underlying data. \citet{2007A&A...471..301U} also suggested that the mean duration for grand minima is 70 years but claimed that the durations of grand minima show a bimodal Gaussian distribution implying two kinds of minima. We also investigated the waiting time distribution for the other three categories (Figure~\ref{fig:MaxMinsub}b, c and d), but were unable to detect any tendencies due to the poor statistics.

Based on the results of the waiting time distributions (Figure \ref{fig:MaxFBMin}b), it is noteworthy that all the grand minima, which are following a grand maximum, occur within 450 years from the peak-time of a grand maximum during the Holocene period. The cumulative probability for a grand minimum to occur as a function of time distance from the peak-time of the Modern Maximum around 1980 is listed in Table~\ref{tab:GrandMaxima}. Under the assumption that the pattern, in which 71\% of grand maxima are followed by a grand minimum, will continue after the termination of the Modern Maximum, the results suggest that there is a 73\% probability that this reduced activity period will reach its lowest values before around 2155 (within 175 years), which is in accordance with the findings of \citet{2010ASPC..428..287A}. However, the overall probability, which is independent of the assumption that 71\% of grand maxima are followed by a grand minimum will continue after the termination of the Modern Maximum, for a reduced activity period to reach its lowest values before around 2155 (within 175 years) is $\sim$0.52.

\begin{table}
\caption{Expected date corresponding to the dip of a grand minimum that follows the Modern Maximum and its probability to occur for assumption dependent and independent cases (see text).}
\begin{tabular}{ccc}
\hline \hline
 Modern Minimum 	& Assumption Dependent 		& Assumption Independent  \\  
 				& Probability				& Probability			 \\ \hline
before 2055		&$\sim$0.33				&	$\sim$0.23		\\
before 2105		&$\sim$0.53				&	$\sim$0.38		\\
before 2155		&$\sim$0.73				& 	$\sim$0.52		\\
\hline
\end{tabular}
\label{tab:GrandMaxima}
\end{table}

\section{The Sun's large-scale dipolar magnetic field over the last three solar cycles}
 
Sunspot observations have showed that Solar cycle 24 is weaker than Solar cycle 23, with peak sunspot numbers almost half the value of Solar cycle 23. Additionally, geomagnetic activity levels during the ascending phase of Solar cycle 24 were lower than any comparable period during Solar cycle 17--23 \citep{2013JSWSC...3A..08R}. Fewer coronal mass ejections have been observed during the ascending phase of Solar cycle 24 and they tend to be less massive and slower when compared to Solar cycle 23 \citep{2014ApJ...784L..27W}. Further, a similar long-term decreasing trend can also be observed in the SMP reconstruction by \citet{2011JGRA..116.2104U}.  All these observations may imply that the current period of reduced magnetic activity have already started during the descending phase of Solar cycle 23. In fact, after the first sunspot maximum observed in 1999, the dynamics of Solar cycle 23 was characterised by a prolonged descending phase with a small-amplitude secondary peak \citep{2004ApJ...601.1136D}. Together with the prolonged descending phase, a significant reduction was observed in the heliospheric magnetic field and the solar wind speed and density during Solar cycle 23 \citep{2011GeoRL..3819106O}. It is thus believed that the Modern Maximum may have ended with Solar cycle 23 in 2000 \citep{2001SoPh..203..179R}.
 
To investigate whether this steady decreasing trend after $\sim$1980 can also be observed in the Sun's large-scale dipolar magnetic field, which is linked to the solar activity proxies such as sunspots and solar modulation potential, we used synoptic photospheric magnetic field maps of the radial magnetic field (B$_{r}$) obtained from line-of-sight magnetogram observations by the Wilcox Solar Observatory (WSO) \citep{1977SoPh...54..353S,2009Hoeksema}. 

The WSO data used in this study span the time period 1976--2014, starting with Carrington rotation (CR) 1642 (27 May 1976), and ending with CR~2155 (14 October 2014). For each map per CR, we carried out spherical harmonic decomposition of the WSO data using the Legendre-transform software provided by the PFSS (Potential Field Source Surface) package of SolarSoft following the recipe given in \citet{2012ApJ...757...96D}. As a result, we obtained the total energy of the large-scale dipolar magnetic field of the Sun. 

\begin{figure}
\resizebox{\hsize}{!}
{\includegraphics{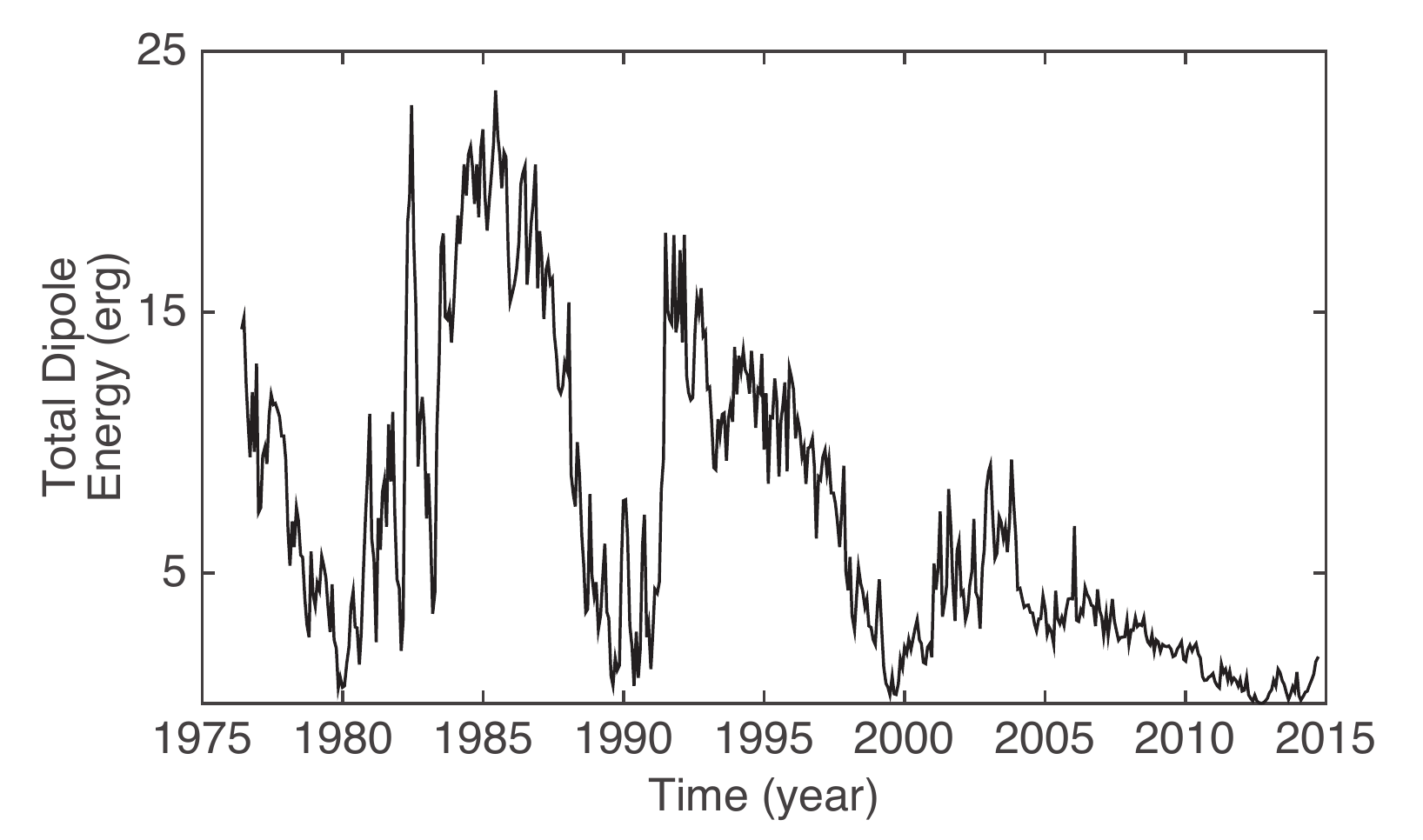}}
\caption{The total energy in the large-scale dipolar magnetic field of the Sun calculated using the Wilcox Solar Observatory data.} 
\label{fig:DipEn}
\end{figure}

The total energy in the Sun's large-scale dipolar magnetic field shows a steady decreasing trend after $\sim$1985 (Figure~\ref{fig:DipEn}) in accordance with the decreasing trends observed in the solar activity proxies \citep{2011JGRA..116.2104U,2011GeoRL..3819106O,2013JSWSC...3A..08R,2014ApJ...784L..27W}. 

\subsection{The $\ell$=1 m=0 and $\ell$=1 $\mid$m$\mid$=1 components of the Sun's large-scale dipolar magnetic field}

The surface magnetic field of the Sun is characterised by the presence of two polar caps, northern and southern solar poles, during the sunspot activity minima. The northern and southern solar polar caps exhibit opposite polarities and they posses the same polarity of the trailing flux within active regions from the same hemisphere emerged during the preceding sunspot cycle. The $\ell$=1 m=0 component of the Sun's large-scale dipolar magnetic field is known to be sensitive to these polar caps and exhibits its maximum energy during sunspot minima. In contrast, $\ell$=1 $\mid$m$\mid$=1 component of the Sun's large-scale dipolar magnetic field attains its maximum energy during the sunspot maxima, when active regions on the photosphere contribute energy to $\ell$=1 $\mid$m$\mid$=1 component, which is oriented in the solar equatorial plane \citep{2012ApJ...757...96D}.

After the termination of the Maunder Minimum, the Sun's large scale magnetic field has stayed largely dipolar, which also behaves quiet unexpectedly since the end of the Modern Maximum \citep{1997A&A...322.1007T,2009SoPh..255..143A}. We thus investigate the $\ell$=1 m=0 and $\ell$=1 $\mid$m$\mid$=1 components of the Sun's large-scale dipolar magnetic field to get a deeper insight.

\begin{figure}
\resizebox{\hsize}{!}
{\includegraphics{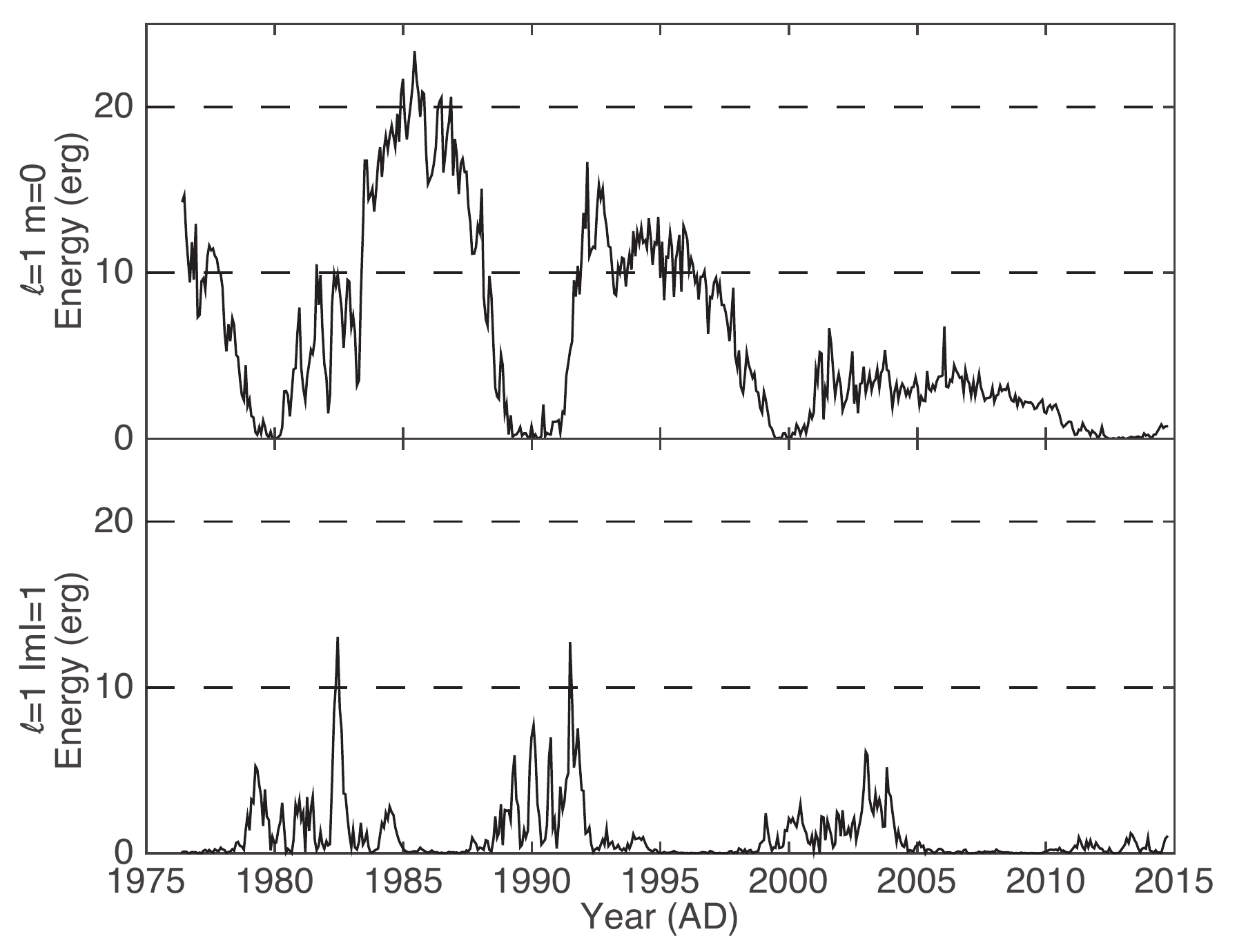}}
\caption{The top panel shows the temporal evolution of the energy in the $\ell$=1 m=0 component of the solar large-scale magnetic field, while the bottom panel represents the temporal evolution of the energy in the $\ell$=1 $\mid$m$\mid$=1 component. From the Wilcox Solar Observatory data} 
\label{fig:AxEqDip}
\end{figure}

Figure~\ref{fig:AxEqDip} shows the $\ell$=1 m=0 and $\ell$=1 $\mid$m$\mid$=1 components of the solar large-scale dipole magnetic field. The energy in the $\ell$=1 m=0 component (top panel of Figure~\ref{fig:AxEqDip}) attains its maximum energy during the sunspot minimum, while the $\ell$=1 $\mid$m$\mid$=1 component of the large-scale magnetic field displays its maximum energy during the sunspot maximum. As seen in the figure, the $\ell$=1 m=0 and $\ell$=1 $\mid$m$\mid$=1 components of the large-scale dipole magnetic field of the Sun has steadily decreased during solar cycles 21--23. 

\begin{figure}
\resizebox{\hsize}{!}
{\includegraphics{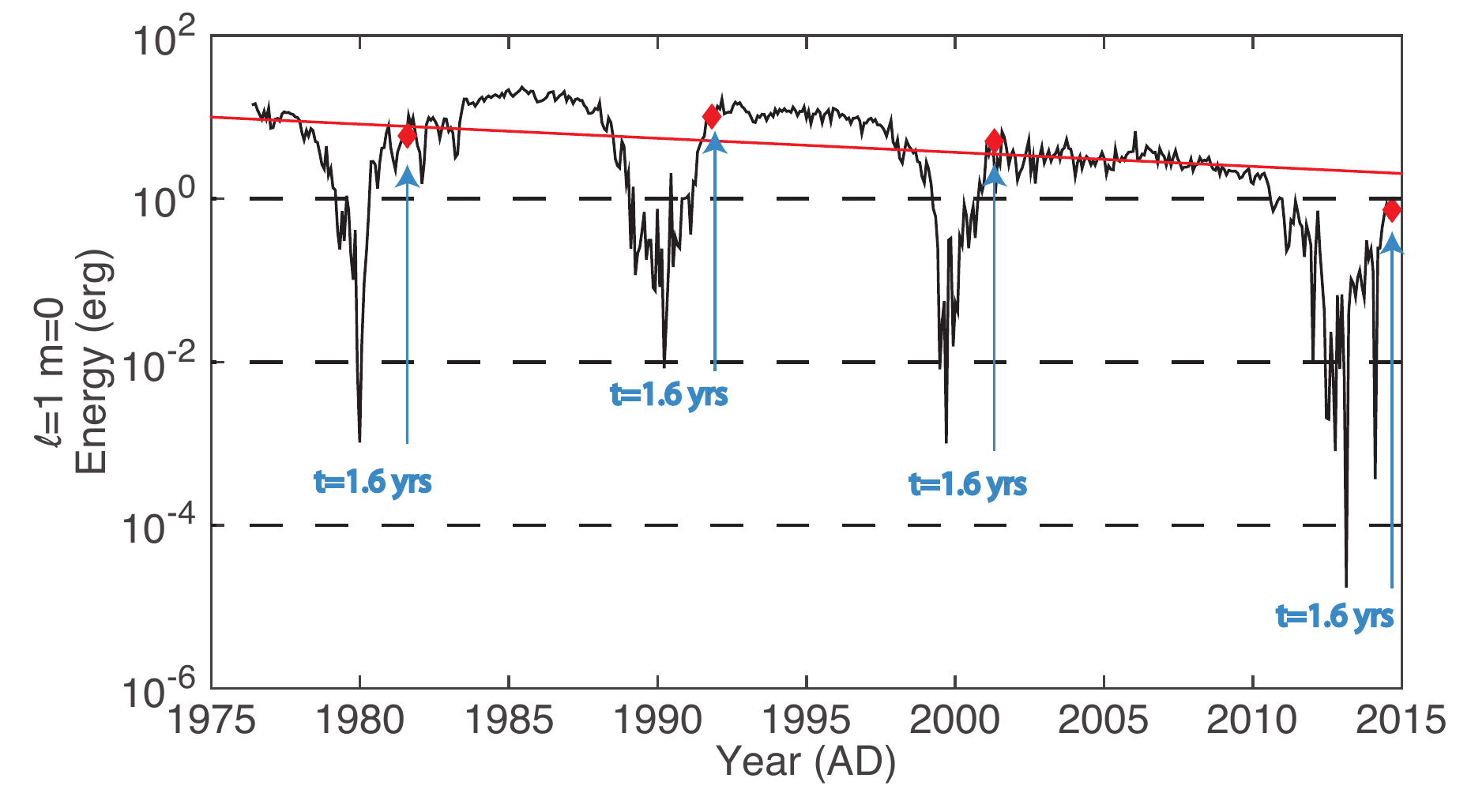}}
\caption{The temporal evolution of the energy in the $\ell$=1 m=0 component of the solar large-scale magnetic field in log-scale. The red diamond show the energy, where 1.6 years have elapsed (blue arrows) since the $\ell$=1 m=0 component reached its lowest value.} 
\label{fig:answer}
\end{figure}

Figure~\ref{fig:answer} shows the temporal evolution of the $\ell$=1 m=0 field in log-scale in order to highlight its behaviour during the maximum phases of solar sunspot cycles 21--24. The deepest minimum of the $\ell$=1 m=0 component over the last three sunspot maxima has been reached during the maximum of solar cycle 24 (Figure~\ref{fig:answer}) and the way it approaches the maximum differs from cycle to cycle. The maximum value reached by the $\ell$=1 m=0 component does not depend on how deep its minimum is, as solar cycle 21 and 23 had almost comparable values during their minima, but their maximum was different. In fact, the strength of the $\ell$=1 m=0 component is a combination of three processes; the total magnetic flux emerging in active regions, the characteristic tilt of this active regions and the meridional flow of the emerging flux and these three different mechanisms are peculiar of each cycle and will determine the strength of the poloidal field of the forthcoming cycle \citep{2013ApJ...767L..25M,2014ApJ...780....5U}. Therefore, what could be an important parameter is how fast the $\ell$=1 m=0 component recover itself from its lowest value. In order to get a measure, we compare the energy in the $\ell$=1 m=0 field to those observed during the previous solar cycles. In particular, at the current solar cycle, 1.6 years had elapsed since the $\ell$=1 m=0 component was at its minimum. We then determined the energy in the $\ell$=1 m=0 component for the previous cycles at the time when 1.6 years have elapsed after they reached their lowest values. The $\ell$=1 m=0 component (red diamonds in Figure~\ref{fig:answer}) had already recovered itself after 1.6 years from its minimum in the previous cycles (Figure~\ref{fig:answer}). This finding indicates that over the same amount of time, the $\ell$=1 m=0 field is still very well below the values of previous cycles during the sunspot maximum of cycle 24. If the strength of the $\ell$=1 m=0 component keeps increasing, then the time to reach its maximum will be longer than that of solar cycle 21, 22 and 23 implying only a slow-down. On the other hand, if the strength does not increase more than the current value, this would imply a weaker solar cycle 25.

\section{Discussion and Conclusion}

The behaviour of solar magnetic activity as seen in solar modulation potential covering the period between 6600 BC--1820 AD has shown that 71\% of identified grand maxima are followed by a grand minimum, revealing an interesting pattern that mainly grand maxima are followed by a grand minimum. This could resemble the behaviour of a relaxation oscillator, which occurs as the result of the dynamical non-linearity induced by the Lorentz force \citep{1995MNRAS.273.1150T,1996A&A...307L..21T,1997A&A...322.1007T}. According to \citet{1975JFM....67..417M}, the dynamo-generated magnetic field will, in general, produce a Lorentz force that will tend to oppose the driving fluid motions through velocity fluctuations, the so-called Malkus-Proctor effect. These fluctuations can be observed on the Sun as torsional oscillations, patterns of zonal flow bands that migrate toward the equator and poles during the solar cycle \citep{1980ApJ...239L..33H}. This mechanism is expected to reduce the amplitude of differential rotation until the effective dynamo falls back to its critical value, at which point the dynamo again saturates. This is an interesting scenario, since a slow down in the differential rotation has already been shown during the Maunder Minimum \citep[][and references therein]{2002SoPh..207..219V}. If the back reaction of the Lorentz force is indeed behind the origin of grand minima, it would also imply that these quiescent periods of activity are not the result of a random process, but instead their origin is linked to the driving mechanism of the magnetic field generation, and therefore their occurrences are not expected to be purely random as shown in the previous studies \citep{2007A&A...471..301U,2015A&A...577A..20I}. This is one of the most important feature that are not reproduced by dynamo models taking into account stochastic noise to generate grand minima \citep{2008SoPh..250..221M}, where the governing parameters of the solar dynamo change randomly.

The back-reaction of the Lorentz force has been shown to lead to a variety of behaviours characterising the long-term solar variability, including amplitude and parity modulation \citep{1995MNRAS.273.1150T,1996A&A...307L..21T,1997A&A...322.1007T}. The latter is of particular interest as strong parity modulation (switching from dipole to quadrupole magnetic field) tend to occur when the Sun is entering or about to leave a grand minimum state. Previous studies suggested that during the Dalton Minimum, the Sun experienced a north-south asymmetry, where sunspots were dominant in the northern solar hemisphere \citep{2009ApJ...700L.154U}, while during the Maunder Minimum sunspots were mainly observed in the southern solar hemisphere \citep{1993A&A...276..549R}. Therefore, we wonder whether or not an increase in the north south asymmetry could possibly be used as a precursor of a grand minimum state \citep{2013ApJ...765..100S,2013ASPC..478..167S}. The current solar dynamics show that even though an excess in sunspot activity in the southern solar hemisphere compared to the northern solar hemisphere has been observed \citep{2009JGRA..114.4101L}, no magnetic confinement in one of the two hemispheres has been detected. According to these results, the Sun's large scale magnetic field is mainly dipolar with a minor contribution of the quadrupole family. Additionally, it is observed that the energy in the large-scale dipolar magnetic field of the Sun (Figure~\ref{fig:DipEn}) and the solar activity indices \citep{2004ApJ...601.1136D,2011GeoRL..3819106O,2014ApJ...784L..27W} show a decreasing trend after the 1980s. This behaviour resembles the Type 2 modulation of the basic 11-year cycle as suggested by \citet{1997A&A...322.1007T}, where large changes in magnetic energy and no changes in parity can be observed and the modulation must therefore result from the action of the velocity perturbations \citep[see][for details]{1997A&A...322.1007T}.

It is still unclear whether the pattern in which $\sim$71\% of grand maxima are followed by a grand minimum, will continue after the termination of the Modern Maximum. The amplitude of the maximum of solar cycle 24 is comparable to the maximum of solar cycle 14 \citep{2014ApJ...780....5U} and confinement of the sunspot activity in one solar hemisphere due to the quadrupole family is not observed and also the value of the $\ell$=1 m=0 field, although weaker, do not yet show a clear drop compared to previous cycles. Our results imply that the Modern Maximum has ended with solar cycle 23 and we do not expect solar cycle 25 to be stronger than solar cycle 24. Therefore, it seems that the Sun has entered a period of moderate activity level and that has $\sim$23\% probability to enter a grand minimum state before 2055.

%%%%%%%%%%%%%%%%%%%%%%%%%%%%%%%%%%%%%%%%%%%%%%%%%%%%%%%%%%%%%%%%%%%%%%%%%%%
\acknowledgments
Funding for the Stellar Astrophysics Centre is provided by the Danish National Research Foundation (Grant agreement no.: DNRF106). FI acknowledge M. DeRosa for providing his code for calculations of the energy in the large-scale dipole magnetic field of the Sun. RS acknowledge the support from the PiCARD collaboration. MFK, CK and JO acknowledge support from the Carlsberg Foundation and Villum Foundation. We would like to thank T. Hoeksema for free access to the WSO information.

%\acknowledgment US spelling: \verb+\acknowledgment+
%\acknowledgement British  spelling: \verb+\acknowledgement+

%%%%%%%%%%%%%%%%%%%%%%%%%%%%%%%%%%%%%%%%%%%%%%%%%%%%%%%%%%%%%%%%%%%%%%%%%%%

\end{document}